# Impact of Information and Communication Technology on Individual Well-being

Aneel Bhusal*

## Abstract:

This paper investigates the impact of information and communication technology (ICT) adoption on individual well-being. Adoption of ICT and its effect on society and individual have been largely mute on the SAARC nations. Using panel data on the 7 membership countries of SAARC, this paper uses Bass diffusion to calculate the coefficient of imitation of these countries for the subscription of each technology in the process of adoption of technology. This degree of imitation has been transformed into the dummy variable based on the average value of the coefficient of imitation for each technology. Econometric model is formed based on the panel data of the SAARC nation to explore the impact of technology subscription on the individual wellbeing of the nation along with other variables. It discusses the various ways how technology like cell phones, telephones and broadband can make markets more efficient and how the diffusion of information and knowledge plays into economic development and growth. Using random effect approaches, the ICT technology variables was found to have a negative and significant impact on countries' level of GDP growth rate. However, when ICT technology when integrated with financial system i.e. the interaction of ICT technology and financial system variable was found to have positive and significant impact on countries' level of GDP growth rate. This study investigates the effect of cell phones, telephone and broadband subscription on the individual wellbeing and growth by performing an econometric analysis on various individual wellbeing variables like life expectancy and expected years of schooling. Overall, the technology subscriptions rate was found to have a positive and significant impact on countries' individual wellbeing.

# 1. Introduction

ICT like telephone, broadband and mobile devices have infiltrated and revolutionized the modern world. Although the effects of these ICT devices and services on society are vast and can be examined through a variety of disciplines, this study will focus on measuring the impact of ICT on individual wellbeing and growth. Through econometric analysis, the study seeks to parse out the direct contribution of the proliferation of ICT on individual development and wellbeing. Furthermore, this investigation will seek to explain the various factors contributing to economic and individual development in order to isolate the true effect of ICT technology. They have the potential to reduce the costs of communication by lowering search costs and making information more accessible to the general population of developing countries. This, in turn, will lead to more efficient market operation by reducing the amount of waste caused by spoilage, and by facilitating communication between producers, sellers, and buyers. ICT use can stimulate the economy by creating more demand for mobile-based services, which in turn increases employment. Mobile phones also offer the potential for mobile phone-based services and products. One example is m-banking, or mobile banking. In this application, users are 2 able to transfer money between bank accounts and pay bills via phone (Akeer & Mbiti, 2014). Cellular mobile, telephone and broadband proliferation in the society will bring about the efficient and positive market contribution and can help the socio- economic growth from various dimension. Although almost anything can be considered technology in one way or another, this chapter focuses on the use of information and communication technologies (ICTs). Information and communication technologies is a broad term that encompasses a variety of communication devices and applications, such as radio, TV, cell phones, computers, computer and network hardware and software, and a variety of applications for these technologies, such as gaming, social networking, instant messaging (IM), and texting. In the past decade, the percentage of young adults using ICTs has increased dramatically. ICTs are now used to fulfill routine tasks such as paying bills and obtaining daily news information; technology diffusion into all aspects of life is ongoing, and despite some barriers (such as access, cost, and skill), the acceptance rates of technology among the population can be high. Internet exploded during the late 1990s into a powerful new social institution" (Goldsmith, 2000). It is now a heavily relied on source of reference material for the public, transcending existing geographical and regulatory boundaries and blurring distinctions between professions and expertise. The use of computers, the Internet, and cell phones are the ICTs discussed in this chapter because they have become mainstreamed into society. Large percentages of individuals, particularly young adults, use these technologies, and public access to and availability of computers and the Internet exists in most communities (Lenhart, Madden, & Hitlin, 2005). According to recent Pew Internet Studies reports, 73 percent of U.S. adults are Internet users, and on average, about 70 million U.S. adults use the Internet on a given day ( (Jack & Jasmine, 2003). Other research shows that communication is the main use of the Internet and generation opportunity of entrepreneurship (Fox, 2004). As of 2005, cell phones represented about 43 percent of all South Asian people have access to one type of communication device (Lipscomb, Apte, & Jemmy, 2006). By investigating the role that played by ICT on GDP growth rate, this study will provide further insight into an existing technology to use it to promote individual wellbeing. Furthermore we can look into the impact of ICT technology into health and education which are other factors of the individual wellbeing. This study will also evaluate how to use existing technology creatively and properly for GDP growth and individual wellbeing.

Section II provides background information theories of individual development and wellbeing along with technology development. It discusses the specific role of information in terms of growth and how ICTs aid the spread of information and knowledge. This section also provides an overview of the existing literature in the area of information technology and individual wellbeing and development, citing both empirical studies and case studies. Section III includes a description of the data and the variables used in the study, as well as the sources from which these data derive. Creation of ICT dummy variable and creation of interaction term are discussed in this section. Section IV explains the modeling approaches used for available panel data while Section V describes the results obtained from each model for each technology. Section VI is a discussion of the overall findings and policy implications with conclusions.

## 2. Literature Review
### 2.1 Economic Theory

The Role of Information in Economic Development Much in accordance with endogenous development hypothesis, information and communication technology affect financial advancement and development essentially through their capacity as a medium of correspondence. They enhance data sharing, which is urgent to the dispersion of thoughts that endogenous development hypothesis accentuates. Data and correspondence innovations decouple data from a "physical store," empowering the spread of data, thoughts, and information that is so basic amid the improvement procedure (Bedia, 1999). The less demanding trade of thoughts can lessen the information hole among created and creating countries, empowering creating nations to expand their ways of life. Data innovations, for example, PDAs, can build efficiencies inside a nation by empowering the trading of data among its occupants and bringing down the 7 cost of obtaining data. Cell phones are particularly critical in creating countries where the necessities of partitioned bunches inside the populace may contrast significantly (Unwin, 2009). Makers and shoppers, most of the populace, would rather require data about business openings, costs of products, training, wellbeing, adequate standards of conduct, and decisions. With mobile phones, unmistakable gatherings can get the specific data they require. The utilization of cell phones additionally suggests a two-way correspondence. After people get the data they require, they can convey their different needs to representing bodies. In this way, mobile phones increment the stream of data, and in addition its general accessibility. Bedia (1999) recommends that in creating nations, dependable data correspondence advances bring down the expenses of transmitting data, which moves the data supply bend to one side. The advances can enhance the nature of data by giving progressive and finish information. With more rich and exact data, individuals in creating nations will have the capacity to settle on better and speedier choices so as to encourage financial development and improvement and diminish destitution (Ronash & Jacob, 1991). Cell phones multiply learning, helping people in the public eye impart and build up a mind boggling system of data. They additionally bring down inquiry costs, diminish the level of awry data in business sectors, and decrease value scattering (Abraham, 2007). ICT technology additionally upgrades the spread of data through system impacts. As an ever increasing number of clients are connected into a data arrange, organize externalities are created in the market, giving an advantage to natives of creating nations and gaining first movers advantage (Bedia, 1999).

## 2.2 Empirical Findings

The growth impact of ICT technology is substantially higher in developing countries than in developed ones. Furthermore, the study provides policy implications for the use of technology to promote global growth (Lum, 2011). The income and government trade policies influence ICT diffusion. However, freedom indices may or may not affect ICT diffusion. Moreover, only personal computers and internet hosts seem to have a positive association with income. Contrary to expectations, ICT diffusion is not associated with education (Baliamoune-Lutz, 2003). China's transformation in new economic growth has been possible by the China's ICT industry development and its diffusion in recent years. Although there is still a huge gap between China and the developed countries in the development of the ICT industry, the astonishing pace of its progress shows promise for the country's New Economy (Meng & Li., 2002). In OECD countries, the international comparison allows relating growth patterns to institutional and policy indicators, thereby offering some preliminary insights into the potential sources of growth disparities. Cross-country evidence yields some tentative support to the idea that institutional factors affecting competition in the product market are likely to affect productivity patterns, especially in a period of rapid diffusion of a general-purpose technology resulting in GDP growth rate (Bassanini &

Scarpetta, 2002). In the study by Du, it was found that ICT helps to promote stock market indirectly positively affecting the GDP (Du, 2018). In the study of Europe, ICT diffusion and ICT usage play a key role and depend on the right framework conditions, not necessarily on the existence of a large ICT-producing sector which increases the market activities (Colecchiaa, 2002). In Brazil and India, the evidence suggests that ICT adoption and productivity has achieve the faster economic growth rate and present country as one of the major economic forces in the world (Basant, Commander, Harrison, & Mene, 2006). The study done in Bangladesh shows that the socioeconomic changes that are caused due to mobile penetration in SAARC countries and found that the changes are in positive direction. So we can say that the increasing rate of mobile penetration will definitely bring socio-economic development in any country. Therefore, importance should be given to enrich this sector by the Government of Bangladesh (Chowdhury, 2015). The mobile cellular subscriptions rate was found to have a positive and significant impact on countries' level of real per capita GDP and GDP growth rate (Chowdhury, 2015).

# 3. Description of Data

The panel data used in this study are drawn from a secondary source such as World Bank and UNDP database. The data used consists of statistics on seven SAARC membership countries over the period 1990 to 2015. The study focuses on this interval because during this time new ICT technology like cell phones and broadband technology first began to come into use and expanded rapidly. The primary variable of interest in this study is the information and communication technology subscription variable. This paper uses three different proxy variables such as the number of mobile cellular, telephone and broadband subscriptions per 100 people. It includes both post-paid and prepaid subscriptions. These subscription variable are converted into the dummy variables based on Bass model estimate of the coefficient of imitation (detailed information is presented in 4.1 section and Appendix). The reason to use degree of imitation is the low level of technological innovation in the sample countries. Other variable used in the research are GDP growth rate, financial system deposits, Gross savings, unemployment growth rate, population growth rate, inflation rate and governmental expenditure. Where GDP growth rate is our dependent variable whereas all other are independent variable.

*Table 1: Descriptive Statistics of Main Dependent and Independent Variables*

| Variable | Observation | Mean | Standard Deviation | Minimum | Maximum |
|---|---|---|---|---|---|
| GDP growth (annual %) | 182 | 5.59 | 3.12 | -8.12 | 9.88 |
| Fixed broadband subscriptions (per 100 people) | 182 | 0.58 | 1.26 | 0 | 6.47 |
| Fixed telephone subscriptions (per 100 people) | 182 | 3.55 | 3.80 | 0.20 | 17.24 |
| Mobile cellular subscriptions (per 100 people) | 182 | 28.42 | 43.25 | 0 | 206.65 |

As we can see minimum and maximum value of fixed broadband subscriptions are 0 and 6.47 which indicates that at the beginning of 1990 the subscribers of the ICT technology were 0 and increasing since 1990 to 2015. However, the less people are using the fixed broadband as the cost price of the broadband is high for the normal people. Similarly, the minimum and maximum value of fixed telephone subscriptions is between 0.20 and 17.24. The cost of operating fixed telephone is high and presence of alternative such as cellular mobile phone people tend to replace telephone. Similarly minimum and maximum value of the mobile cellular subscriptions is between 0 and 206.65. The minimum value is zero because of less presence of mobile in 1990s and after that the mobile subscriptions have increased rapidly. The maximum value of cellular mobile subscription is 206.65 indicating that one people in this region has almost two mobile cellular in 2015. Since, cellular mobile can replace function of broadband and telephone the subscription rate of mobile phone is high compared to the broadband and telephone. The minimum and maximum value of GDP was -8.1247394 and 9.88 respectively.

### 3.1. Bass Model of Diffusion (Creation of ICT Dummy variables)

We use Bass Model of Diffusion (Bass, 1969) in order to estimate the number of technology adopted by users of each technology per 100 people. Using Bass Model of Diffusion, we estimate the diffusion of subscriber of each technology per 100 people. We calculate the coefficient of innovation and the coefficient of imitation for each country based on technology subscription. These subscription variables are converted into the dummy variables based on Bass model estimate of the coefficient of imitation. The reason to use degree of imitation is the low level of technological innovation in the sample countries. We use degree of imitation for each country as the determining factor for the dummy variable on the use of the technology.

The Bass Model of Diffusion equation is given as

$$\frac{f(t)}{1-F(t)} = p + qF(t) \quad \ldots\ldots(A)$$

Where:

$f(t)$ is the change of the installed base fraction

$F(t)$ is the installed base fraction

$p$ is the coefficient of innovation

$q$ is the coefficient of imitation

We use the subscription of ICT technology per 100 people to estimate the number of technology adopted per 100 people. The result of the Bass Model of Subscription is shown in the Appendix section.

Dummy variable is created for broadband, telephone and cellular technology from the Bass Model of diffusion's degree of imitation of each technology. The average value of degree of imitation for the sample countries was calculated. The sample countries with above the average of the degree of imitation was considered with the dummy as 1 and the country with below the average of the degree of imitation was considered with the dummy as 0 for each ICT subscription variable broadband, telephone and cellular.

# 4. Empirical Methodology:
## 4.1 Random Effects Approach

In order look the impact of dummy variable of ICT subscription, we use random effects approach. Since panel data were available for this study, a random effects approach was tested in order to account for other variables over time and by country that may affect GDP growth rate of the country. For each i, the equation was averaged over time to obtain:

$$gdpgrowthrate = \beta 0 + \beta 1. CellularDummy + \beta 2. FinancialSystemDepositstimesCellularDummy + \beta 3. FinancialDepositstoGDP + \beta 4. GrossSavings + \beta 5. Unemploymentrate + \beta 6. PopulationGrowthRate + \beta 7. Inflationrate + \beta 7. Expense + \alpha i + \mu t \ldots\ldots\ldots\ldots\ldots(B)$$

Random effects estimation is useful in this case because there might be an unobserved effect, αi that affects a country's economic growth. The willingness of a country to adopt new technology, for example, may impact growth. The dependent variable in equation is the GDP growth rate.

*FinancialSystemDepositstimesCellularDummy* variable is the interaction term created by multiplying *CellularDummy* variable and *FinancialDepositstoGDP* variable. As financial system has improved by the introduction of the ICT, this interaction term should have the positive relation with the GDP growth rate. *CellularPhoneDummy* is created based on the Bass model estimate of the degree of imitation which should have positive relation with GDP growth rate. *FinancialDepositstoGDP* is the deposits made by the people in the financial system. The financial deposits to GDP should have positive relation with GDP growth rate. *GrossSavings* is the savings made by the economy. As the gross savings brings the investment, the *GrossSavings* should have positive relation with GDP growth rate. *UnemploymentRate* brings problem in the economy and damages the economy therefore *UnemploymentRate* should have negative relation with GDP growth rate. *InflationRate* increases costs of production and damages the economy, therefore InflationRate should have negative relation with GDP growth rate. *Expense* is the expenditure made by the government in the economy, therefore it should have positive relation with GDP growth rate.

# 5. Results

## 5.1 Random Effects Approach:

The result of Random Model in GDP growth rate is given in the Table 1.

*Table 2: Random Effects Model: Dependent Variable: GDP Growth Rate*

|                              | (1) Cellular | (2) Telephone | (3) Broadband |
|---|---|---|---|
| Cellular Dummy               | -5.6668*** |  |  |
|                              | (1.626)    |  |  |
| Financial system*Cellular    | 0.1283***  |  |  |
|                              | (0.0404)   |  |  |
| Financial deposits to GDP (%)| 0.003671   | -0.01928  | -0.009686 |
|                              | (0.0258)   | (0.0286)  | (0.0339)  |
| Gross savings (% of GDP)     | 0.01929    | 0.05151   | 0.04178   |
|                              | (0.0385)   | (0.0385)  | (0.0387)  |
| Unemployment, total (%)      | 0.04771    | 0.1202    | 0.02566   |
|                              | (0.0936)   | (0.117)   | (0.101)   |
| Population growth (%)        | 1.1495***  | 0.9793*** | 0.8211*** |
|                              | (0.273)    | (0.267)   | (0.285)   |
| Inflation (Annual %)         | -0.01898   | -0.02696  | -0.02157  |
|                              | (0.0496)   | (0.0502)  | (0.0506)  |
| Expense (% of GDP)           | 0.008159   | 0.002411  | 0.01551   |
|                              | (0.0439)   | (0.0519)  | (0.0525)  |
| Telephone Dummy              |            | -4.0699*** |  |
|                              |            | (1.505)   |  |
| Financial system * Telephone |            | 0.09917*** |  |
|                              |            | (0.0375)  |  |
| Broadband Dummy              |            |           | -1.3965   |
|                              |            |           | (1.468)   |
| Financial system*Broadband   |            |           | 0.05339   |
|                              |            |           | (0.0382)  |
| Constant                     | 2.9195*    | 3.1802*   | 3.3450*   |
|                              | (1.557)    | (1.624)   | (1.805)   |
| Observations                 | 182        | 182       | 182       |

Standard errors in parentheses
$^* p < 0.10$, $^{**} p < 0.05$, $^{***} p < 0.01$

The result from the above table shows that the main independent variable cellular mobile subscription dummy variable has negative impact on the annual GDP growth rate and is statistically significant. Similarly the telephone and broadband subscription dummy variable also have negative impact on the annual GDP growth rate. While the telephone subscription dummy variable is statistically significant, the broadband subscription dummy variable is statistically insignificant. The slope of ICT dummy variable is negative which tell us that adoption or imitation of ICT technology alone will have negative impact on GDP growth rate. Adoption of ICT technology alone brings the financial leakage in the country's economy and brings about the negative impact in the economic development of the poor country. Only adopting and imitation of ICT technology will increase the expense of the country to acquire such technology (Baliamoune-Lutz M. , 2003). According to The Kathmandu Post, Nepal leading newspaper, Nepal spent Rs 3.7b on bandwidth in the year 2012 (Shrestha, 2013). Nepal can be an indicator for the other SAARC nations. Therefore, only adopting and imitating of ICT technology alone will have negative impact on the GDP growth rate of the country.

Financial deposits to GDP variable is positive but not statistically significant in GDP growth rate for cellular mobile model. However, the financial deposits to GDP is negative in GDP growth rate for broadband and telephone model which are also insignificant. Thus, financial deposits shows a mixed results with GDP growth rate.

The interaction term of cellular subscription dummy variable with financial deposit is positive and statistically significant. Similarly the interaction term of telephone and cellular dummy variable with financial deposit is also positive and is statistically significant. Interaction term is positive slope which tells us that only imitating ICT technology is not sufficient. There needs to be better financial system. The imitation of technology has to be collaborated with the financial system in the different sector in order to increase the GDP growth rate. Thus, GDP growth rate increases when collaboration of ICT technology with the financial system of the country. This is consistent with how the financial system operates in developing countries. For example banking and financial services have improved and become easy because of the implementation of ICT technology.

Therefore, a lot of deposits have been brought to the economy. The introduction of the ICT services like e-banking and mobile banking has made the banking deposits and transaction even easier. According NDTV news, India leading news provider, between year 2016 and 2017 India sees 55%increase in the digital transaction and 122% in mobile transaction (Dutta, 2017). Therefore, implementing ICT technology with the deposits in the financial system will increase the money in the economy and will bring about investment in the economy. Therefore only adopting and imitating of ICT technology is not sufficient for GDP growth. Countries need better financial system which can be collaborated with ICT technology for the positive impact of ICT technology on the GDP growth rate of the country.

The result is fairly robust, and the impacts of the additional explanatory variables were similar across all the models. Other control variables have mixed results. Population growth in particular has interesting impact as it contributes to GDP growth rate positively and is statistically significant. This is consistent with the results from other countries. For example the population growth rate in USA has positive impact in the GDP growth rate of the country (Kitov , 2008). Financial system deposits has positive impact on the GDP growth rate as the money in the system increases (Beck, Demirgüç-Kunt, & Levine, 2000). Gross savings in the system has a positive impact in the system (Jappelli & Pagano, 1994). Unemployment rate brings about the positive relation with the GDP growth rate. This totally contradicts with unemployment brings negative impact on the GDP growth rate (Iwuamadi & Awogbenle, 2010). Expense made in the economy brings about the positive impact on the economic development (Barron, 1990). Thus, when implementing ICT policy as developing indicator in a country. Country should try to implement and imitate ICT technology with collaboration of financial systems of the country to bring about positive GDP growth rate in a country.

Previous table analyzes the impact of ICT on GDP growth rate for the financial wellbeing of the individual. But the goal of this paper is to analyze the impact of ICT on other aspects of individual well being too. So the following section analyzes the impact of ICT on other measures of individual wellbeing which is life expectancy and expected year of schooling.

### 5.1.1 Robustness Check with Alternate Dependent Variables

The result of Random Model on Life Expectancy is given in the Table 2.

*Table 3: Random Effects Model: Dependent Variable: Life Expectancy*

|  | (1) Cellular | (2) Telephone | (3) Broadband |
|---|---|---|---|
| Cellular Dummy | -0.5537 | | |
|  | (2.377) | | |
| Financial system*Cellular | -0.07765 | | |
|  | (0.0561) | | |
| Financial deposits to GDP (%) | 0.1448*** | 0.1071*** | 0.1058*** |
|  | (0.0344) | (0.0372) | (0.0403) |
| Population growth (%) | 0.07449 | -0.5967* | 0.6118* |
|  | (0.379) | (0.351) | (0.348) |
| Gross savings (% of GDP) | 0.2080*** | 0.2036*** | 0.2568*** |
|  | (0.0503) | (0.0502) | (0.0455) |
| Unemployment, total (%) | 0.8114*** | 0.5101*** | 0.4090*** |
|  | (0.135) | (0.151) | (0.138) |
| Inflation (annual %) | -0.08326 | -0.08986 | -0.1399** |
|  | (0.0653) | (0.0653) | (0.0607) |
| Expense (% of GDP) | 0.2423*** | 0.2029*** | 0.5662*** |
|  | (0.0780) | (0.0759) | (0.0852) |
| Health expenditure(% of GDP) | 0.2278 | 1.3177*** | 0.1715 |
|  | (0.242) | (0.252) | (0.204) |
| GDP per capita growth (%) | 0.1661* | 0.2974*** | 0.2432*** |
|  | (0.101) | (0.0990) | (0.0884) |
| Telephone Dummy | | 9.1756*** | |
|  | | (2.153) | |
| Financial system * telephone | | -0.1428*** | |
|  | | (0.0503) | |
| Broadband Dummy | | | -3.9386** |
|  | | | (1.721) |
| Financial system*Broadband | | | -0.04014 |
|  | | | (0.0450) |
| Constant | 47.008*** | 42.956*** | 44.858*** |
|  | (2.074) | (2.192) | (2.129) |
| Observations | 182 | 182 | 182 |

Standard errors in parentheses
* $p < 0.10$, ** $p < 0.05$, *** $p < 0.01$

The result from the above table shows that the main independent variable cellular mobile subscription dummy variable has negative impact on the life expectancy of the people but is statistically significant. Similarly the broadband subscription dummy variable also has negative impact on the life expectancy of the people but is statistically significant. However telephone subscription dummy variable has positive impact on the life expectancy of the people and is statistically significant. ICT technology will makes ease of access of the health information and makes it easier to provide emergency services. ICT technology has positive impact on the health services (Reinhold & Jurgen, 2008). This contradicts with the cellular and broadband dummy but is consistent with telephone dummy variable. Therefore, ICT technology telephone has positive impact on the life expectancy.

The interaction term of cellular subscription dummy variable with financial deposit is negative and statistically significant. Similarly the interaction term of telephone and cellular dummy variable with financial deposit is also negative and statistically significant. Since government emphasis its investment in ICT technology development, the expenditure on the health facilities is getting low. Therefore, the interaction term has negative impact on the life expectancy.

The result is fairly robust, and the impacts of the additional explanatory variables were similar across all the models. Other control variables have mixed results. Financial deposits to GDP (%) variable in particular has interesting impact as it contributes to life expectancy of people positively. This result is consistent with result from other countries like in the study done in USA (Scott, Sivey, Ait, & Willwnberg, 2010).

Previous table analyzes the impact of ICT on life expectancy which is important factor for the individual. Again the paper analyzes the impact of ICT on expected year of schooling which is the other measures of individual wellbeing.

The result of Random Model on Expected Year of Schooling is given in the Table 3.

*Table 4: Random Effects Model: Dependent Variable: Expected Year of Schooling*

|  | (1) Cellular | (2) Telephone | (3) Broadband |
|---|---|---|---|
| Cellular Dummy | -4.8593*** | | |
|  | (1.003) | | |
| Financial system*Cellular | 0.05555** | | |
|  | (0.0259) | | |
| Unemployment, total (%) | 0.3018*** | 0.3172*** | 0.1801*** |
|  | (0.0524) | (0.0671) | (0.0595) |
| GDP per capita growth (%) | 0.1123** | 0.2040*** | 0.2261*** |
|  | (0.0461) | (0.0516) | (0.0505) |
| Inflation (annual %) | -0.006855 | -0.02731 | -0.04174 |
|  | (0.0292) | (0.0331) | (0.0330) |
| Population growth (%) | -0.2113 | -0.6291*** | -0.4255** |
|  | (0.158) | (0.175) | (0.184) |
| Gross national expenditure (%) | 0.004744 | 0.02514 | 0.03647** |
|  | (0.0174) | (0.0187) | (0.0176) |
| Gross savings (% of GDP) | -0.04104* | 0.001027 | -0.02441 |
|  | (0.0230) | (0.0272) | (0.0262) |
| Financial deposits to GDP (%) | 0.1077*** | 0.09015*** | 0.05473** |
|  | (0.0168) | (0.0202) | (0.0223) |
| Telephone Dummy | | -0.3356 | |
|  | | (1.026) | |
| Financial system * Telephone | | -0.02791 | |
|  | | (0.0267) | |
| Broadband Dummy | | | -2.6510*** |
|  | | | (0.941) |
| Financial system*Broadband | | | 0.04604* |
|  | | | (0.0248) |
| Constant | 5.5225*** | 3.2148* | 4.0438** |
|  | (1.612) | (1.766) | (1.892) |
| Observations | 182 | 182 | 182 |

Standard errors in parentheses * $p < 0.10$, ** $p < 0.05$, *** $p < 0.01$

The result from the above table shows that the main independent variable cellular mobile subscription dummy variable has negative impact on the expected years of schooling and is statistically significant. Similarly the broadband subscription dummy variable also has negative impact on the expected years of schooling is statistically significant. Telephone subscription dummy variable has negative impact on the expected years of schooling but is statistically insignificant. As more information are available by the introduction of ICT in the educational system, the ICT technology has negative impact on the expected years of schooling. As students can get more knowledge in an easier and faster way, then tend to spend less time in the school to graduate (Pelgrum, 2001). Therefore ICT technology has negative impact on the expected years of schooling as student tend to graduate early.

The interaction term of cellular subscription dummy variable with financial deposit is positive and statistically significant. Similarly the interaction term of broadband dummy variable with financial deposit is also positive and statistically significant. However, the interaction term of telephone dummy variable with financial deposit is negative and insignificant.

The result is fairly robust, and the impacts of the additional explanatory variables were similar across all the models. Other control variables have mixed results. Financial deposits to GDP (%) variable in particular has interesting impact as it contributes to the expected year of schooling positively. Similarly, population growth rate variable in particular has interesting impact as it contributes to expected year of schooling negatively.

The results in Table 2 and Table 3 shows similar results in terms of GDP growth rate, which takes factors such as life expectancy and expected year of schooling into account. The results are fairly robust, and the impacts of the additional explanatory variables are similar across the models.

# 6. Conclusions and Summary

Overall, the results of the study suggest that the growth in ICT technology use has increased rapidly over the past two decades has had a significant effect on GDP growth rate. In the Random Effects Model performed, the ICT dummy variable has had negative and significant impact on the GDP growth rate for a given country. Similarly, interaction term between ICT dummy and financial system deposit variable have a positive slope showing positive impact. The result suggests that only imitating ICT technology alone in personal basis will have negative impact on the GDP growth rate. However, if we are able to integrate the ICT technology with financial sector by the use of ICT technology then the impact is positive. Several statistical models were tested and refined.

The result provides a convincing argument about the benefits generated by technology can be achieved by focusing on the engagement of technology in the financial sector in the SAARC nations. The GDP growth rate of SAARC countries can be increased if ICT technology can be implemented with financial sector. ICT technology exhibited an important and significant impact on the GDP growth rates. Again, the ways that cell phones, telephones and broadband subscription can impact economic development and growth are numerous. Other factors of economic variable like savings, GDP per capita, etc. were taken into account in the study.

Information and communication technology use can reduce search costs and increase information availability, which makes markets function more efficient. In terms of the diffusion of ideas and knowledge, cellular phones, telephones and broadband makes available information in market prices and creates employment opportunities. Services like mobile banking eliminates the need for clients to spend time traveling to the physical banks. Information and communication technology can also be used to deliver important information about health and to increase literacy. The growth of the cell phone, telephone and broadband industry itself has added much to the economy financially, adding more jobs and creating more demand for products and services is another way in which these technology have contributed to economic growth.

Similar results were discovered in terms of level of economic development, which takes factors such as education and life expectancy into account. The results proved fairly robust, and the impacts of the additional explanatory variables were similar across the models. In the study, several statistical models were tested and refined. Random effects models were tested, and instrumental variables were used to control for the possible endogeneity of ICT technology adoption. Information and communication have lately found exceptional use in banking, enabling greater access to capital, which facilitates investment and productivity with adoption. Therefore, the benefit that can be achieved from the adoption of the ICT technology is enormous. Therefore, governments can and should promote the use of ICT services to improve market functionality and the quality of life in SAARC countries. They could accomplish positive GDP growth as promoting use of ICT in financial sector. Governments could also provide subsidies and try to attract foreign investors in the industries. Adopting of ICT technology by country increases the confidence of country investment for rational foreign investor to invest. Furthermore analysis can be done using the data of the subscription model in the random effect model and study the relevant result.

# APPENDIX:

*Table 5 Bass Model of 7 SAARC countries cellular mobile subscription*

| Bass Model of 7 SAARC Countries Cellular Mobile Subscription |||||||||||||
|---|---|---|---|---|---|---|---|---|---|---|---|---|
| | M | | | | P | | | | Q | | | | |
| Country | Estimate | Std. Err. | t-test | p-value | Estimate | Std. Err. | t-test | p-value | Estimate | Std. Err. | t-test | p-value | Dummy | R-squared |
| Banglades | 90.7 | 3.09 | 29.36 | 0 | 7.97E-05 | 3.03E-05 | 2.63 | 0.016 | 0.457 | 0.027 | 16.95 | 0 | 0 | 0.99584 |
| Bhutan | 85.46773 | 2.83159 | 30.18 | 0 | 0.01303 | 0.00321 | 4.06 | 0.0023 | 0.54743 | 0.06043 | 9.06 | 0 | 0 | 0.9928 |
| India | 77.1 | 1.63 | 47.31 | 0 | 2.12E-05 | 1.43E-05 | 1.48 | 0.16 | 0.726 | 0.0546 | 13.3 | 0 | 1 | 0.9943 |
| LKA | 105 | 2.12 | 49.72 | 0 | 1.24E-05 | 7.06E-06 | 1.76 | 0.091 | 0.572 | 0.0354 | 16.14 | 0 | 0 | 0.99508 |
| Maldeivs | 192 | 5.8 | 33.2 | 0 | 0.00207 | 0.00074 | 2.8 | 0.012 | 0.464 | 0.0432 | 10.8 | 0 | 0 | 0.99015 |
| Nepal | 110 | 4.97 | 22.12 | 0 | 0.000458 | 0.000133 | 3.45 | 0.0039 | 0.519 | 0.033 | 15.71 | 0 | 0 | 0.99672 |
| Pakistan | 67.4 | 1.16 | 58.33 | 0 | 1.38E-07 | 1.66E-07 | 0.83 | 0.41 | 0.877 | 0.0731 | 12 | 0 | 1 | 0.99352 |
| | | | | | | | | | Mean=0.594633 | | | | | |

*Table 6 Bass Model of 7 SAARC Countries Telephone Subscription*

| Bass Model of 7 SAARC Countries Telephone Subscription |||||||||||||
|---|---|---|---|---|---|---|---|---|---|---|---|---|
| | M | | | | P | | | | Q | | | | |
| Country | Estimate | Std. Err. | t-test | p-value | Estimate | Std. Err. | t-test | p-value | Estimate | Std. Err. | t-test | p-value | Dummy | R-squared |
| Banglades | 3.92346 | 0.18606 | 21.09 | 0 | 0.01168 | 0.00972 | 1.2 | 0.2421 | 0.19117 | 0.14764 | 3.13 | 0.0049 | 0 | 0.85295 |
| Bhutan | 0.7475 | 0.0693 | 10.78 | 0 | 0.0396 | 0.021 | 1.89 | 0.072 | 0.1976 | 0.1269 | 1.56 | 0.134 | 0 | 0.76948 |
| India | 3.1726 | 0.1847 | 17.18 | 0 | 0.0278 | 0.0304 | 0.92 | 0.369 | 0.4074 | 0.2797 | 1.81 | 0.083 | 1 | 0.68126 |
| LKA | 16.28413 | 1.15224 | 14.13 | 0 | 0.00113 | 0.00118 | 0.95 | 0.35006 | 0.41123 | 0.09763 | 4.21 | 0.00036 | 1 | 0.90938 |
| Maldeivs | 9.268 | 0.625 | 14.82 | 0 | 0.0164 | 0.128 | 1.28 | 0.21 | 0.184 | 0.399 | 0.46 | 0.65 | 0 | 0.49927 |
| Nepal | 3.38161 | 0.20137 | 16.79 | 0 | 0.0127 | 0.00332 | 3.83 | 0.00092 | 0.32264 | 0.04197 | 5.3 | 0 | 1 | 0.9654 |
| Pakistan | 3.2438 | 0.5274 | 6.15 | 0 | 0.1577 | 0.0484 | 3.26 | 0.0036 | 0.4882 | 0.1895 | 1.62 | 0.076 | 1 | 0.74853 |
| | | | | | | | | | Mean = 0.314606 | | | | | |

*Table 7 Bass Model of 7 SAARC countries for Broadband Subscription*

| Bass Model of 7 SAARC Countries for Broadband Subscription |||||||||||||
|---|---|---|---|---|---|---|---|---|---|---|---|---|
| | M | | | | P | | | | Q | | | | |
| Country | Estimate | Std. Err. | t-test | p-value | Estimate | Std. Err. | t-test | p-value | Estimate | Std. Err. | t-test | p-value | Dummy | R-squared |
| Bangladesh | 4.19673 | 0.39732 | 10.56 | 0 | 0.005387 | 0.00787 | 6.84 | 0.001 | 0.44282 | 0.10316 | 4.29 | 0.0077 | 0 | 0.99446 |
| Bhutan | 7.745193 | 4.994965 | 1.55 | 0.17 | 0.00069 | 0.00025 | 2.84 | 0.0297 | 0.725468 | 0.153047 | 4.74 | 0.003 | 1 | 0.9081 |
| India | 13.12213 | 0.23587 | 55.6 | 0 | 0.000156 | 0.00042 | 3.7 | 0.003 | 0.684144 | 0.040412 | 16.9 | 0 | 1 | 0.9705 |
| LKA | 35.93451 | 0.451061 | 7.97 | 0 | 0.00239 | 0.00086 | 2.77 | 0.017 | 0.434661 | 0.063102 | 6.89 | 0 | 0 | 0.98727 |
| Maldeivs | 5.67933 | 0.17765 | 31.97 | 0 | 0.00661 | 0.00425 | 1.56 | 0.1478 | 0.86846 | 0.15021 | 5.78 | 0.0001 | 1 | 0.97761 |
| Nepal | 1.05 | 0.0648 | 16.15 | 0 | 7.5E-05 | 0.00018 | 0.41 | 0.6939 | 0.16 | 0.045 | 3.56 | 0.009 | 0 | 0.9690 |
| Pakistan | 10.1819 | 0.0357 | 28.52 | 0 | 0.000266 | 0.00145 | 1.84 | 0.1 | 0.92801 | 0.11452 | 8.1 | 0 | 1 | 0.99135 |
| | | | | | | | | | Mean=0.80623 | | | | | |